\documentclass[letterpaper,14pt]{article}
\usepackage[centertags]{amsmath}
\usepackage{amsfonts}
\usepackage{amssymb}
\usepackage[english]{babel}
\usepackage[dvips]{graphicx}
\begin{document}\begin{center}
{Semi-classical approach to quantum black holes }
\end{center}
\begin{center}
{Euro Spallucci\footnote{Corresponding author e-mail: euro@ts.infn.it} and Anais Smailagic}
\end{center}

\begin{center}Department of Theoretical Physics,
         University of Trieste, Strada Costiera 11, 34014 Trieste,
         Italy, and Sezione INFN di Trieste,
         Strada Costiera 11, 34014 Trieste, Italy
\end{center}
\vskip 1cm

\begin{center}
\begin{abstract}

Semi-classical description of BH, as it was originally introduced by Hawking and Bekenstein in the early seventies, where
classical solutions of the Einstein equations are coupled to quantum matter fields, opened for the first time a  window with
a glance on the quantum aspects of gravity. The surprising properties showed by semi-classical description of black holes (BHs) 
called for a full quantum  description for such objects. In particular, the understanding  of BH entropy in terms of micro-states,
as well as the behavior during the final phase of evaporation, cannot be properly described in semi-classical terms.

Quantum gravity has a long story of doomed attempts to cure the theory from non-renormalizability and quantum anomalies.
In the eighties Green and Schwarz brilliantly
solved these problems in the framework of $SO(32)$ or $E_8 \times E_8$ super-symmetric quantum string theories. Up to now, string theory 
is the best candidate for a grand unified theory including gravity at the quantum level in a consistent way. Accepting that the
building blocks of matter are one-dimensional extended objects, characterized by a length scale of order $10^{-33} cm$, still 
 the relation between fundamental strings and BHs remains to be determined. The exponential increase of string states
degeneracy with the string excitation energy resembles the exponential increase of BH states with the increase of
BH mass. This and other results suggest a correspondence between BHs and highly excited strings, or string
balls. However, a more precise stringy formulation of BHs requires a non-perturbative formulation of the theory itself,
including the interaction with higher dimensional objects, i.e. D-branes. So far, this approach provides the correct entropy
counting BH micro-states only in a limited number of cases, but not in general.

In this Chapter we would like to review a "~phenomenological~" approach taking into account the most fundamental feature of
string theory or, more in general, of quantum gravity \cite{kiefer}, whatever its origin, which is the existence of a minimal 
length in the space-time fabric \cite{book1}. This length is generally identified with the Planck length, 
or the string length, \cite{Fontanini:2005ik,Aurilia:2013mca} but it could be also 
much longer down to the TeV region. A simple and effective way to keep track of the effects the minimal length in BH 
geometries is to solve the Einstein equations with an energy momentum tensor describing non point-like matter. The immediate consequence
is the absence of any curvature singularity. Where textbook solutions of the Einstein equations loose any physical meaning because
of infinite tidal forces, we find a de Sitter vacuum core of high, but finite, energy density and pressure. An additional improvement
regards the final stage of the BH evaporation leading to a vanishing Hawking temperature even in the neutral, non-rotating, case.
In spite of th simplicity of this model we are able to describe the final stage of the BH evaporation, resulting in a
cold remnant with a degenerate, extremal, horizon of radius of the order of the minimal length.
In this chapter we shall
describe only neutral, spherically symmetric, regular BHs although charged, rotating  and higher dimensional BHs
can be found in the literature \cite{Smailagic:2010nv,Nicolini:2008aj}
\end{abstract}
\end{center}

\section{Introduction}
The description of radiating BHs by Hawking \cite{Hawking:1974sw} offered the first physically relevant ``~peep~'' on the mysteries
of quantum gravity. After more than forty years of intensive research in this field ( see \cite{BD,Fulling,PD} and \cite{Hollands:2014eia} for a recent review with an  extensive reference list ) various aspects of the problem still remain 
to be properly explained. 
For example, a satisfactory description of the terminal stage of BH evaporation remains still to be understood. 
So far, the string theory seems to be the best candidate for self-consistent, ultraviolet completion of gravity 
at the Planck scale. On the other hand, following Bekenstein's idea, the BH entropy is 
formally identified with the area of
the event horizon in Planck units. Thus, combining Hawking's definition of BH temperature with the Area Law gives
a consistent thermodynamical description of semi-classical BHs dynamics. However, to give a physical
ground to this interpretation   one needs a proper identification of BH micro-states, i.e. a complete statistical interpretation
of entropy. Beforehand one can say that the number of micro-states has to grow exponentially with the area, but the physical 
meaning of these micro-states is elusive in this semi-classical quantum framework. Contrary to the classical gas thermodynamics,
where the entropy is a function of the volume enclosing the system, in the BH case entropy is a function of the 
area of the object indicating that the eventual BH micro-states should lie somewhere on the horizon surface instead of in the internal
volume \footnote{This is a possible realization of the Holographic Principle which frequently pops up when dealing 
with quantum gravity  \cite{Susskind:1994vu,Susskind:1998dq}. 
It seem to be a general property of self-gravitating quantum system that physical
degrees of freedom  are confined on the boundary enclosing the system, rather than in the bulk.  } \\
Interestingly enough, in string theory the degeneracy of excitation states also shows the same exponential growth. 
This suggests an identification among highly excited string states and BH micro-states. 
However, a more precise stringy formulation of Bhs requires a non-perturbative formulation of string theory itself,
including the interaction with higher dimensional objects, i.e. D-branes. So far, this approach provides the correct entropy
counting BH micro-states at least in a limited number of cases, but not yet in general.\\

In this Chapter we would like to review a "~phenomenological~" approach taking into account the most fundamental feature of
string theory or, more in general, of quantum gravity, whatever its origin, which is the existence of a \emph{minimal length} in
the space-time fabric \cite{book1}. This length is, so far, identified with the Planck length, or the string length, but nothing
prevents it from being much longer, even  down to the TeV region. A simple and effective way to keep track of the modifications
of BH geometries, due to the presence of the minimal length, is to solve the Einstein equations with an energy momentum tensor 
describing non point-like matter. \\
At first glance,one could think of modifying the four dimensional Einstein action
to incorporate minimal length effects. To solve the modified field equations can be quite complicated. This is the case, where
minimal length is introduced in the Einstein theory through the, so-called, ``star-product'', i.e. the standard product
between fields is replaced by a non-commutative product operation making even the simplest field theory non-local and
impossible to solve. The only way to handle non-local field theories is in a truncated perturbative expansion suppressing
non-locality and introducing spurious derivative interactions. In the resulting approximate theory, the effect of the minimal
length as a natural short-distance cut-off is lost and paradoxically infrared divergences appear even in the massive case
and mix with the persisting ultraviolet divergences.  A typical example of a ``... cure killing the patient!''.\\ 
The star-product approach, while formally correct is of little practical use. For example, BH perturbative solutions  
of ``starred-product'' Einstein equations, manifest the same pathological short distance behavior of the corresponding
standard solutions, in spite of introducing a short-distance cut-off \cite{Nasseri:2005ji,Nasseri:2005yr}.\\
We argued that an alternative approach is to implement
the minimal length \emph{only} in the matter source. The line of reasoning is the following. The metric field is a geometrical 
structure defined over an underlying manifold and the curvature is a measure of the strength of the metric field. Thus, 
it is the response to the presence of a mass-energy distribution. The minimal length is an 
intrinsic property of the manifold itself, rather
than a super-imposed geometrical structure. It affects gravity in a subtle, indirect way: it
smears matter  eliminating point-like objects as physical sources of the gravitational field. As a matter of fact,
this idea goes hand-in-hand  with the fundamental idea of string theory: ``particles'' are not matter-points, rather they
are extended, one dimensional  objects, whose  length is of the order of the Planck length, or any other fundamental length scale.
Following the above reasoning, we conclude that in General Relativity
the effects of the minimal length can be taken into account by \emph{keeping the standard form} of the Einstein tensor 
in the l.h.s. of the field equations and introducing a \emph{modified energy-momentum tensor} as a source in the r.h.s.\\
At this point one may object that the textbook approach defines BHs as ``matter-free'' solutions of the field equations.
Even in the case of the Reissner-Nordstr\"om BH, the energy momentum tensor \emph{only} describes the electric field.\\
This way of introducing BHs is, at least, misleading and contradicts the fundamental ideas of General Relativity that
the geometry of space-time is determined by the energy-momentum distribution.The \emph{only} globally defined
 matter-free solution of the Einstein equations is Minkowski space-time. Furthermore, solving ``in vacuum'' faces the problem
of \emph{a posteriori} determination of various integration constants. A crystal clear example of this difficulty
is the Kerr solution, where integration constants are fixed by comparison with the geometry of a slowly rotating
sphere, while everybody is aware that the rotating sphere is \emph{not} the source of the the Kerr gravitation field! \\
Furthermore,  the symmetry of the ``vacuum'' solution,  has to be imposed \emph{a priori}, since there is noting
in the field equations that can do the job for you. \\
All the above mentioned ambiguities disappear once a proper matter source is introduced in the Einstein equations.

\section{``~Renormalizing~'' BHs: a toy-model}
\label{toy}
A well known example, in quantum field theory, of difficulties related to modeling elementary particles as structureless
point like objects, is the appearance of ultraviolet infinities. 
When computing one-loop Feynman graphs, one finds infinite, unphysical, amplitudes for measurable processes.
The simplest ``cure'' of the infinity disease is to introduce a suitable short-distance cut-off, which in coordinate-space
is simply a ``minimal length''.  Once divergent parts of the amplitudes are subtracted away, the (arbitrary) cut-off 
is replaced by sum physically meaningful quantity through a ``re-normalization'', i.e. re-parametrization, of the theory.\\
Following this philosophy, we propose to re-normalize the curvature singularity in a BH geometry in a similar fashion.\\
First, let us cut-off the singularity by introducing a minimal length into the line element.\\
Second, relate the minimal length to the radius of the smallest physically meaningful,\emph{extremal} BH  .\\
This two-step procedure will be first applied to the simples case of Schwarzschild BH. In order to remove the curvature 
singularity in $r=0$, we introduce the simplest kind of short-distance cutoff through the replacement

\begin{equation} 
 \frac{1}{r}\longrightarrow \frac{r^2}{r^3 + l_0^3}
\label{uno}
\end{equation}

where, $l_0$ is, for the moment, an arbitrary length scale. According with the framework one has in mind, $l_0$ can be thought 
of as the characteristic length scale of the underlying quantum theory of gravity (whatever it is), i.e. the Planck length, 
the string length, etc.  In the technically much more involved  approach of ``asymptotically safe'' quantum gravity, 
 a similar ``cutoff identification'' was  introduced in \cite{Bonanno:2000ep} for the same purpose.   
\footnote{ Alternative choices of the cutoff function are

\begin{equation}
 \frac{1}{r}\longrightarrow \frac{r^2}{r^3 + \omega G_N \left(\, r + \gamma G_N M\right)}\nonumber
\end{equation}
where, $\omega $ and $\gamma$ are constants coming from non-perturbative renormalization group 
calculations \cite{Bonanno:2000ep}; or \cite{Hayward:2005gi}
\begin{equation}
 \frac{1}{r}\longrightarrow \frac{r^2}{r^3 + 2M G_N  l_0^2 }\nonumber
\end{equation}
}, 
By skipping all the
approach-dependent technicalities we  write a regularized Schwarzschild metric as

\begin{equation}
 ds^2= -\left(\, 1 - \frac{2G_N \mathcal{M}\left(\, r\,\right)}{r}\, \right) dt^2 
+\left(\, 1 - \frac{2G_N M\left(\, r\,\right)}{r}\, \right)^{-1} dr^2 + r^2 d\Omega^2
\label{rbh}
\end{equation}

where

\begin{equation}
 \mathcal{M}\left(\, r\,\right)\equiv M \frac{r^3}{r^3 + l_0^3/2}
\label{massar}
\end{equation}

From the equation (\ref{massar}) one sees the $ \mathcal{M}(r) \longrightarrow M $ at large distance, i.e. $r>> l_0$, 
while in the opposite 
limit $ M(r) \longrightarrow 2M r^3/l_0^3 $. As a consequence the metric approaches the standard form of the Schwarzschild 
line element at large distance, and the deSitter metric

\begin{equation}
 ds^2= -\left(\, 1 - \frac{4 G_N M }{l_0^3} r^2\, \right) dt^2 
+\left(\, 1 - \frac{4G_N M }{l_0^3} r^2\, \right)^{-1} dr^2 + r^2 d\Omega^2
\end{equation}

at short distance, with an effective cosmological  constant $\Lambda= 12 G_N M/l_0^3 $. The deSitter geometry
is known to be curvature singularity-free. As a matter of fact, we have replaced a physically meaningless infinite
curvature point with a  central deSitter condensate, where the curvature can be very high but finite. 
The physical mechanism leading to the disappearance of the singularity is clear, as the deSitter vacuum exerts a \emph{negative}
pressure balancing the gravitation pull towards the center. Thus, close to the origin, collapsed matter is supported by
the negative pressure of the deSitter vacuum and can attain an equilibrium configuration. \\
Step two, requires to identify the physical meaning of $l_0$ through an in depth study of the BH geometry.\\
The eventual horizons are obtained from the equation \footnote{ The radius of the inner and outer horizon can be obtained
by solving the cubic algebraic equation

\begin{equation}
r_H^3 -2MG_N r_H^2 +\frac{l_0^3}{2}=0 
\end{equation}
For the sake of completeness, we give the exact solutions:

\begin{eqnarray}
&& r_{<0}=\frac{2MG_N}{3}\left(\, 1-2\cos\theta\,\right)\ ,\nonumber\\
&& r_-=\frac{2MG_N}{3}\left(\, 1+2\cos\left(\,\theta-\frac{\pi}{3}\,\right)\,\right)\ ,\nonumber\\
&& r_+=\frac{2MG_N}{3}\left(\, 1+2\cos\left(\,\theta+\frac{\pi}{3}\,\right)\,\right)\ ,\nonumber\\
&&\cos(3\theta)=-1+2\left(\, \frac{3l_0}{4MG_N}\right)^3\nonumber
 \end{eqnarray}

The first root is an unphysical negative solution. $r_\pm$ are the radii of the inner/outer horizons.
%\begin{equation}
% r_+ \approx 2MG_N \left(\, 1 - \frac{l_0^3}{32 M^3 G_N^3}\,\right)
%\end{equation}
}  :

\begin{equation}
 M= \frac{r^3_H+ l_0^3/2}{2G_N r^2_H}
\label{horeq0}
\end{equation}

\begin{figure}[ht!]
\begin{center}
\includegraphics[width=8cm,angle=0]{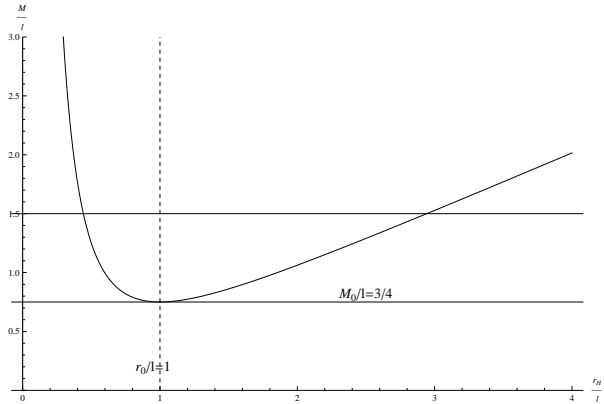}
\caption{\label{mass} Plot of the function in equation (\ref{horeq0}) }
\end{center}
\end{figure}

The function (\ref{horeq}) has a single minimum in 

\begin{eqnarray}
 && r_0= l_0\ ,\\
 && M_0= \frac{3l_0}{4G_N}
\end{eqnarray}

describing an ``\emph{extremal}'' BH in the sense that for any $M> M_0 $, the radius of the outer horizon, $ r_+$, 
is larger than $l_0$.  

%One of the most identifications  of $l_0$ is with the Planck distance $l_P\equiv \sqrt{G_N}= l_0$.
%In this case, the extremal BH is a Planckian object with
%
%\begin{eqnarray}
% && r_0= l_P\ ,\\
% && M_0= \frac{3M_P}{4}
%\end{eqnarray}

The extremal BH can be understood as the thermodynamical equivalent of the ``minimal volume'' of a real gas $V_0$.
As it known, the entropy of the gas is of the form $S = k_B \ln (V/V_0)$ leading to $S=0$ for $V=V_0$. In the BH case, the role
of the gas with volume $V_0$ is played by  the extremal BH which is expected to have  zero entropy 
in agreement with the third law of thermodynamics.\\

A second, inner horizon  $r_-$, exists for any non-extremal configuration (see Fig.(\ref{mass})). $r_+$ and $r_-$ merge 
into $r_0$ for $M\to M_0$. For $M<M_0$ no BH exists. Thus, $l_0$ has been promoted from an arbitrary parameter with length 
dimension, to radius  of the horizon of the minimal size extremal BH  \footnote{If the smallest object
can be produced is an extremal BH of size $l_0$ \cite{Mureika:2011hg,Nicolini:2013ega,Spallucci:2011rn}, 
there is no physical way to probe sub-Planckian distances \cite{Dvali:2010bf,Dvali:2010jz,Nicolini:2012fy}. }.\\
Contrary to the case of  the ``bare'' Schwarzschild BH,  the renormalized metric
 admits and extremal configuration even in the absence of electric charge and angular momentum.\\ 
 
The existence of a de Sitter region at short distance, indicates the presence of a non-trivial energy-momentum tensor sourcing
the renormalized metric (\ref{rbh}). This energy-momentum tensor must approach a cosmological form at short-distance.  
The energy density for a spherically symmetric mass distribution is given by

\begin{equation}
 \rho\equiv \frac{1}{4\pi r^2} \frac{d\mathcal{M}}{dr}= \frac{3M}{8\pi}\frac{l_0^3}{\left(\, r^3 +l_0^3/2\,\right)^2}
\end{equation}

Let us remark that the invariant energy dentity is finite everywhere. In particular 
$\rho \left(\, 0 \,\right) = 3M/2\pi l_0^3  $ and this is a necessary condition to avoid the
appearance of a curvature singularity  in the origin. \\
The complete energy momentum tensor encodes energy and pressure distributions of an anisotropic fluid  
\cite{Nicolini:2005vd} given by

\begin{equation}
T^\mu{}_\nu = -p_\theta \delta^\mu{}_\nu + \left(\, \rho +p_\theta\,\right) \left(\,- u^\mu u_\nu +l^\mu l_\nu\,\right)
\label{source}
\end{equation}

where, $u^\mu= \left(\, 1\ , 0\ , 0\ , 0\,\right)$, $l^\mu= \left(\, 0\ , 1\ , 0\ , 0\,\right)$, 
and $p_\theta= \rho + (r/2)d\rho/dr $
follows from the energy-momentum tensor covariant divergence-free condition. $p_\theta$ is finite everywhere as well.
Thus, we have a regular source leading to the everywhere finite  curvature geometry described by the line element (\ref{rbh}).

\subsection{Thermodynamics }

The thermodynamical description of BH starts with the computation  of the Hawking temperature defined as the surface gravity
of the Killing horizon over $4\pi$: 

\begin{equation}
 T_H\equiv \frac{\kappa_H}{4\pi}=\frac{1}{\sqrt{-g_{00}g_{rr}}}\left\vert\frac{dg_{00}}{dr}\right\vert_{r=r_+}
\label{th}
\end{equation}

In our case, $-g_{00}g_{rr}=1$ and (\ref{th}) leads to

\begin{equation}
 T_H=\frac{1}{4\pi r_+}\frac{r_+^3- l_0^3}{ r_+^3+ l_0^3/2 }
\end{equation}

First, we notice that $T_H$ vanishes at the extremal configuration, $r_+=l_0$, which eliminates the pathological behavior
of the Schwarzschild solution at the final phase of the evaporation. $T_H$ increases with increasing radius up to a maximum
value at $r_{max.}= (5+3\sqrt3)^{1/3}l_0/2^{1/3}$, then approaches zero as $r_+\to \infty$.\\

\begin{figure}[ht!]
\begin{center}
\includegraphics[width=8cm,angle=0]{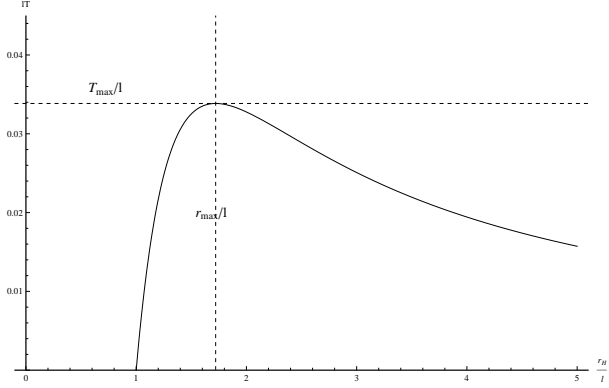}
\caption{\label{temp} Plot of the Hawking temperature  as a function of the horizon radius.}
\end{center}
\end{figure}

Another important thermodynamical quantity assigned to the BH is its entropy. After the Bekenstein/Hawking results, it is
customary to identify the entropy with $1/4$ of the horizon area in Planck units. Suppose we have a Planckian BH with radius 
and mass given by 

\begin{eqnarray}
 && r_0= l_P\ ,\\
 && M_0= \frac{3M_P}{4}
\end{eqnarray}

where, $l_P$ and $M_P$ are the Planck length and mass. Continuing the previous discussion,  the extremal Bhs  correspond to
 the molecular volume. In a statistical description the extremal BH should correspond to a macro-state realized by a 
\emph{single} micro-state and, thus, has zero entropy.\\
Apart from the general assumption of the validity of the area law, one should be able to calculate it in the same way 
as it is calculated for a real gas, i.e. from the first law of thermodynamics. In the case of a BH, the first law is given by:

\begin{equation}
 dM= T_H dS \label{1law}
\end{equation}
 
From (\ref{1law}) one can derive that

\begin{equation}
 dS= 4\pi \left[\, \frac{\partial g_{00}}{\partial M}\, \right]^{-1}_{r=r_+} dr_+
\label{ds}
\end{equation}

For the metric (\ref{rbh}) one finds 

\begin{eqnarray}
 S &&= \frac{\pi}{G_N}\left(\, r_+^2 - l_0^2\,\right) +\frac{\pi}{G_N}l_0^3 \left(\, \frac{1}{l_0} -\frac{1}{r_+}\,\right)\ ,
\nonumber\\
   &&= \frac{A_H}{4G_N}\left(\, 1 -\frac{V_0}{V_H}\,\right)
\label{entropy}
\end{eqnarray}

where,
\begin{eqnarray}
&& A_H\equiv 4\pi r_+^2\ ,\\
 && V_H \equiv \frac{4\pi}{3} r_+^3\ ,\\
&& V_0\equiv \frac{4\pi}{3}  l_0^3
\end{eqnarray}

For large BHs, far away from extremality, $V_H>> V_0$, we recover the standard area law $S=A_H/4G_N$, but as the ``quantum'' 
regime is approached volume corrections become important and cannot be neglected.\\
It is crucial to remark that the integration of (\ref{ds}), leading to (\ref{entropy}), is bounded from below by the
radius $r_0$ of the extremal configuration as there are no smaller BHs.\\
Thus, we obtain the generalized ``~area law~'' which also gives the  form of the entropy satisfying the third law. It contains further
corrections induced by the presence of the minimal length. For the ``bare'' Schwarzschild metric there is no extremal, minimal size, configuration
and the corresponding entropy is simply on fourth of the area. Assuming that this is a ``universal'' behavior for all
BHs, leads to an inconsistency with the third law, i.e. zero-temperature extremal BHs with non-zero entropy. Thus, even in the 
absence of a minimal length, whenever a BH admits an extremal configuration, the correct way to calculate the entropy is by integrating
the first law from the minimal radius of the extremal configuration and not from zero. The correct definition
of entropy is given by the difference of the area of the non-extremal configuration and the extremal one. 

\subsection{Phase transitions}

In this subsection we are going to study eventual phase transitions of regular neutral BHs. 
We follow the analogy with finite temperature quantum field theory where different vacuum
phases are studied in terms of the stationary points of effective potential. The order parameter is the vacuum
expectation value of some scalar field operator expressed as a function of the temperature. Different phases correspond
to different vacuum expectation values.   \\
In our case  off-shell free energy, $ F^{Off.} $ plays the role of the effective potential while the order parameter 
is represented by the radius of the BH. Different phases correspond to different size BHs.
The advantage of using  the \emph{off-shell free energy} is that $T$ is a free parameter  describing the evolution from
non-equilibrium, $T\ne T_H$,  to $T=T_H$ states when the BH is in equilibrium with the surrounding thermal bath.

\begin{equation}
 F^{Off.}\equiv M - T S
\end{equation}

We look for the extremal of $ F^{Off.} $ which corresponds to equilibrium configurations.

\begin{equation}
 F^{Off.}=\frac{1}{4G_N r_+^2}\left(\, 2r_+^3 + l_0^3 - 4\pi r_+^4 T + 4\pi l_0^3 r_+ T\, \right)\ , r_+\ge l_0
\label{fofff}
\end{equation}
The extrema of (\ref{fofff}) are solutions of the condition  

\begin{equation}
 \frac{dF^{Off.}}{dr_+}=\frac{1}{2G_N r_+^3}\left(\, r_+^3 - l_0^3 - 4\pi r_+^4 T - 2\pi l_0^3 r_+ T\, \right)=0
\label{fprimo}
\end{equation}

Equation (\ref{fprimo}) determines the free parameter $T$ as a function of $r_+$

\begin{equation}
 T=\frac{1}{4\pi r_+}\frac{r_+^3- l_0^3}{ r_+^3+ l_0^3/2 }\equiv T_H
\label{teq}
\end{equation}

proving the general property that the extrema of $F^{Off.}$ corresponds to BHs in thermal equilibrium with
the surrounding heat bath.\\
If we were able to invert (\ref{teq}) we could find the way in which the order parameter $r_+$ evolves by varying $T$, 
as in the finite temperature quantum field theory. Unfortunately, the equation is fourth order and cannot be easily
and transparently solved. However, we can obtain simple analytic solutions by considering the near-extremal and large radius 
limits for low temperature BHs. \\
For near extremal configurations we find

\begin{equation}
 r_{min.}\approx l_0 \left(\, 1 + 2\pi l_0 T\, \right)
\end{equation}

while, for large BHs, away from extremality, we find 

\begin{equation}
 r_{max.}\approx \frac{1}{4\pi T}
\end{equation}

At low temperature $r_{min.}$ and $r_{max.}$ are the  local minimum and local maximum  of $F^{Off.}$. Thus, near-extremal
BHs are classically stable, while large BHs are unstable and decay either towards extremality or grow indefinitely without
ever reaching the equilibrium.\\  
The two extrema merge at a critical temperature $\tilde{T}$ where both (\ref{fprimo}) and

\begin{equation}
 \frac{d^2F^{Off.}}{dr_+^2}=\frac{1}{2G_N r_+^4}\left(\, 3l_0^3 - 4\pi r_+^4 T +4\pi l_0^3 r_+ T\, \right)
\label{fsecondo}
\end{equation}
vanish. One finds that this happens for

\begin{equation}
 \tilde{r}_{flex.}^3= \frac{l_0^3}{2}\left(\, 5 + 3\sqrt{3}\,\right)=r^3_{max.}
\end{equation}
and 
\begin{equation}
 \tilde{T}=\frac{1}{2\pi \tilde{r}_{flex.}}\frac{1}{1+\sqrt{3}}=T_H^{max.}
\end{equation}

\begin{figure}[ht!]
\begin{center}
\includegraphics[width=8cm,angle=0]{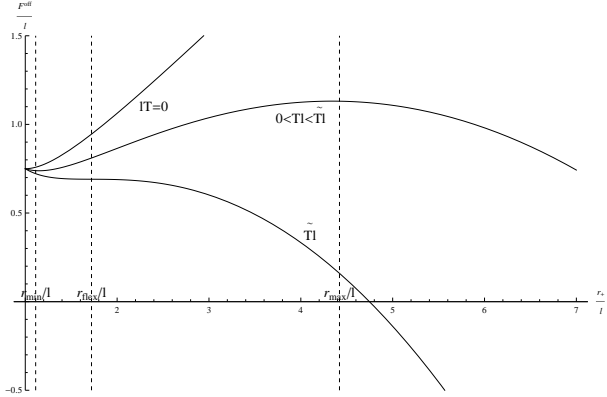}
\caption{\label{free} Plot of the off-shell free energy for different values of the temperature $T$.  }
\end{center}
\end{figure}

Figure (\ref{free}) summarizes the main conclusions of the previous analysis of phase transitions. At zero temperature there is
a single extremal BH in equilibrium with the surrounding vacuum. As the temperature increases, even slightly, the free energy 
develops a local minimum and a local maximum, the first corresponds to a small, near-extremal, classically stable BH, while the 
second is a large classically unstable BH. In fact, the free energy is unbounded from below for large BHs which are colder than
the surrounding heat bath. As a consequence they continue to grow never reaching an equilibrium configuration. 
This behavior follows from the ``competition'' between the internal energy end entropy entering  $F^{Off.}$ with
opposite sign.  $M$ increases linearly
with $r_+$ while the entropy increases with $r_+^2$. Thus, the negative entropy term dominate over the positive internal energy
contribution beyond some critical size. This pathological 
behavior will be cured in the next section by introducing an Anti de Sitter background whose negative (inward) pressure will
stop this unlimited expansion. \\
A further increase in $T$ leads to a critical value, $\tilde{T}$, where the maximum and the minimum merge into an inflexion point.
Remarkably, $ \tilde{T}= T_H^{max.}$, where only one unstable BH exists.\\
For $T> \tilde{T}$ BHs do not exists.\\
The critical behavior of the system can also be inferred from the form of the heat capacity:

\begin{equation}
 C_H\equiv \frac{\partial M}{\partial T_H}= -\frac{3\pi l_0^3}{G_N}\frac{r_+^2 \left(\, r_+^3 +l_0^3/2\,\right)}
{\left[ r_+^3 -l_0^3 \left(\, 5+3\sqrt{3}\,\right) \right]\left[ r_+^3 -l_0^3 \left(\, 5-3\sqrt{3}\,\right) \right]}
\end{equation}

$C_H$ diverges for $r_+ = r_{max.}$, it is positive for $l_0 \ge  r_+ \ge r_{max.}$, and is negative for $r_+ > r_{max.}$.
The near-extremal region has $C_H>0$ and BHs show the normal, stable, thermodynamical behavior, while in the region $C_H< 0$ the anomalous 
behavior described above takes place, i.e. increasing the total mass $M$ lowers the temperature triggering a limitless growth.

\subsection{Area quantization}

The results obtained in the previous discussion indicate that there is a minimal size (extremal) BH even of Planckian dimension, as well
as the corresponding minimal area.  Thus, we are led to an interesting conjecture, i.e. an  holographically improved  ``Bohr quantization'' 
of the BH. Instead of quantizing the mass of the BH, we rather  quantize
its area in terms of the minimal area of the extremal configuration

\begin{equation}
 A_H= n A_0 = 4\pi n\, l_0^2\ ,\qquad n=  1\ , 2\ , 3\ ,\dots
\label{an}
\end{equation}

It follows from (\ref{an}) that the radius of the horizon increases according with 

\begin{equation}
 r_H= \sqrt{n}\, l_0
\label{rn}
\end{equation}

By inserting (\ref{rn}) into (\ref{horeq}) we obtain the quantized mass spectrum as :

\begin{equation}
 M_n= \frac{n^{3/2}+ 1/2}{2G_N n }l_0= \frac{2}{3}M_0 \frac{n^{3/2}+1/2 }{n}
\label{mspectr}
\end{equation}

For large $n$. highly excited  BH states have mass given by

\begin{equation}
 M_n\approx   \frac{2}{3}M_0\, n^{1/2}
\label{mspectr2}
\end{equation}

and the difference between successive mass levels vanishes as $n^{-1/2}$. This is the region where the thermal picture for the 
Hawking radiation makes sense since the mass levels become practically continuous.\\
On the other hand, in the truly quantum regime transitions occur discontinuously through emission of single quanta $\hbar \omega $ 
given by the mass difference between nearby levels:

\begin{equation}
 \Delta M_n = \frac{2}{3}M_0 \left[\, \left(\, n+1 \,\right)^{1/2} -n^{1/2}\,\right]\equiv\hbar \omega_n
\end{equation}

In this picture, the final stage of BH decay resembles more the discontinuous spectra of the atomic transitions, than a thermal
radiation from an hot body.

\section{AdS Black hole and criticality}
The unboundedness from below of the free energy, we found in the previous section, is specific to the asymptotically flat boundary
conditions satisfied by the the metric (\ref{rbh}). To cure the BH instability towards a limitless growth one introduces a negative
cosmological constant in the Einstein equations and solve them with the energy-momentum tensor (\ref{source}).\\
From a physical point of view, a negative cosmological constant represents a positive (inward pushing) vacuum pressure
 
\begin{equation}
 p\equiv -\frac{\Lambda}{8\pi G_N}=\frac{3}{8\pi G_N a^2}
\end{equation}

making an unbounded inflation energetically disfavored.\\
A further motivation for studying AdS BHs is that in higher dimensions they have a pivotal role in the implementation
of the  $AdS_5/CFT$
duality \cite{Maldacena:1997re,Witten:1998zw,Witten:1998qj}. This kind of duality
offers a powerful tool to tackle  non-perturbative features
of a variety of physical systems ranging from the quark-gluon plasma
\cite{Myers:2008fv}
to fluids \cite{Ambrosetti:2008mt} and super-conductors \cite{Hartnoll:2009sz}.
The strong-coupling regime of a conformal field theory living
on the flat boundary of $AdS_5$ is mapped by duality into the weak-coupling
quantum string theory (quantum gravity) in the $AdS_5\times S_5$ bulk. 
This amazing spin-off of string theory connects $4D$ physics in 
flat space-time to quantum gravity in $AdS_5$ and provides a beautiful realization
of the Holographic Principle \cite{Susskind:1994vu,Susskind:1998dq}.\\
Sticking to our toy-model, we find the line element 

\begin{equation}
 ds^2= -\left(\, 1 - 2G_N\frac{ M\left(\, r\,\right)}{r}+ \frac{r^2}{a^2}\, \right) dt^2 
+\left(\, 1 - 2G_N\frac{ M\left(\, r\,\right)}{r} + \frac{r^2}{a^2} \, \right)^{-1} dr^2 + r^2 d\Omega^2
\label{radsbh}
\end{equation}

where $\Lambda\equiv -3/a^2$ and

 \begin{equation}
 M(r)\equiv  M\frac{r^3_+}{  r^3_+ + l_0^3\left(\, 1 + 3l_0^2/a^2\,\right)/2  }
\end{equation}
\begin{figure}[ht!]
\begin{center}
\includegraphics[width=8cm,angle=0]{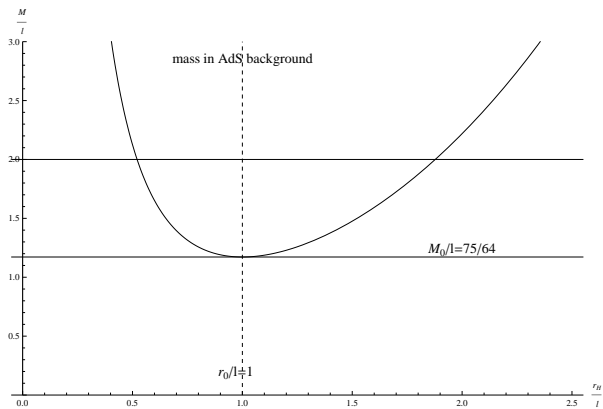}
\caption{\label{mad} Plot of the function in (\ref{horeq}).  }
\end{center}
\end{figure}

We have introduced a new free parameter in the model, i.e. the vacuum pressure, which will play, together
with the temperature, an important role in determining the phases of the system.\\
The two length scale $l_0$ and $a$ rule the short and large distance behavior of the metric, respectively.
The short distance form of the metric is again of the type
\begin{equation}
 ds^2= -\left(\, 1 - \frac{\Lambda }{3} r^2\, \right) dt^2 
+\left(\, 1 - \frac{\Lambda }{3} r^2\, \right)^{-1} dr^2 + r^2 d\Omega^2
\end{equation}
 
where

\begin{equation}
 \frac{\Lambda}{3}=\frac{2G_N M}{l_0^3\left(\, 1 + 3l_0^2/a^2\,\right)/2} -\frac{1}{a^2}
\end{equation}

As the three parameters $M$, $l_0$, $a$ are free, $\Lambda$ can be positive, negative or even zero.  In any case the metric
is singularity free being either deSitter, Anti deSitter,  or Minkowski.\\
The existence of horizons can be seen plotting the BH mass given by

\begin{equation}
 M= \frac{r^3_+ +l_0^3\left(\, 1 +3 l_0^2/a^2 \,  \,\right)  /2}{2G_N r^2_+}\left(\, 1 + \frac{r_+^2}{a^2} \,\right)
\label{horeq}
\end{equation}

as a function of $r_+$. The plot is given in figure (\ref{mad})

\begin{equation}
 \frac{\partial M}{\partial r_+}= \frac{1}{2G_N r_+^3}\left[\, r_+^3 \left(\, 1 + 3 r_+^2/a^2 \,\right)    
-l_0^3\left(\, 1 +3 l_0^2/a^2 \,\right) \,\right]
\end{equation}

The temperature is given as

\begin{equation}
 T_H=\frac{1}{4\pi r_+}
\frac{ r_+^3 \left(\, 1 +3 r_+^2 /a^2\,\right)   - l_0^3 \left(\, 1 +3 l_0^2 /a^2  \,\right) }
{r_+^3 + l_0^3 \left(\, 1 +3 l_0^2/a^2  \,\right)/2}
\label{tadsbh}
\end{equation}

\begin{figure}[ht!]
\begin{center}
\includegraphics[width=8cm,angle=0]{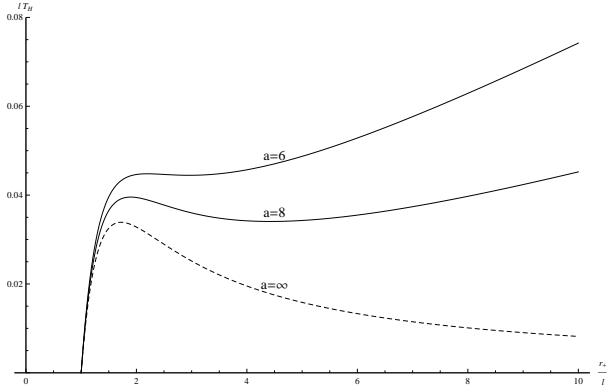}
\caption{\label{tempads}  Plot of the Hawking temperature for different values of the vacuum pressure.
The inflexion point  marks the transition between low/high pressure regimes. }
\end{center}
\end{figure}

The entropy turns out to be 

\begin{eqnarray}
 S&&= \frac{\pi}{G_N}\left(\, r_+^2 - l_0^2\,\right) +\frac{\pi}{G_N}l_0^3 \left(\, 1 +\frac{3 l_0^2}{a^2} \,\right) 
\left(\, \frac{1}{l_0} -\frac{1}{r_+}\,\right)\ ,\nonumber\\
  &&=\frac{A_H}{4G_N}\left(\, 1 -\frac{V_0}{V_H}\,\right)+ \frac{3\pi}{G_N} \frac{l_0^4}{a^2}\left(\, 1 -\frac{l_0}{r_+} \,\right) 
\label{entropy2}
\end{eqnarray}

\begin{eqnarray}
 F^{Off.}&&=\frac{1}{4G_N r_+^2}\left[\, 2r_+^3 + l_0^3\left(\, 1 + 3l_0^2/a^2\,\right) \,\right]
\left(\,  1+ \frac{r_+^2}{a^2}\,\right) \nonumber\\
&&-  \frac{\pi T }{G_N}\left[\, r_+^2 - l_0^2 + 
l_0^3 \left(\, 1 +\frac{3 l_0^2}{a^2} \,\right) \left(\, \frac{1}{l_0} -\frac{1}{r_+} \,\right)\,\right]
\label{foff}
\end{eqnarray}

\begin{figure}[ht!]
\begin{center}
\includegraphics[width=8cm,angle=0]{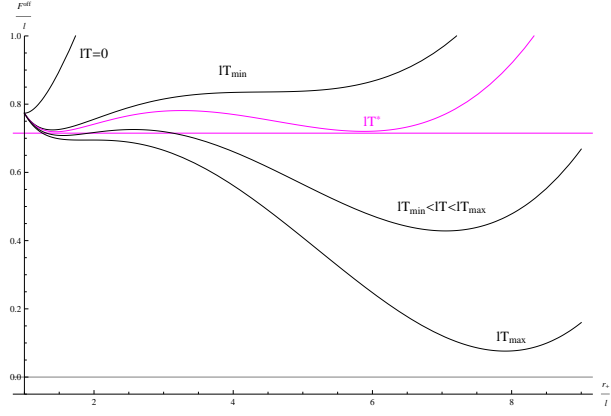}
\caption{\label{fasi}  Plot of the off-shell free energy for different values of the temperature $T$ in the low-pressure regime
showing the existence of different phases. The critical curve, for $T=T^\ast$ is shown in magenta.      }
\end{center}
\end{figure}
Let us note that (\ref{foff}) for large $r_+$ can be approximated with

\begin{equation}
F^{Off.}\approx \frac{r_+^3}{8G_N a^2}-  \frac{\pi T }{G_N} r_+^2
\label{large}
\end{equation}

Equation (\ref{large}) shows how large radius configurations are energetically disfavored as they imply a net increase
of the free energy. The negative increase from the  entropy term  is no more able to compensate for positive increase
of internal energy.

\begin{eqnarray}
 \frac{dF^{Off.}}{dr_+}&&=\frac{1}{2G_N r_+^3}\left[\,  r_+^3 +3 r_+^5/a^2 -l_0^3\left(\, 1 + 3l_0^2/a^2\,\right)\right.\nonumber\\
&& - 4\pi r_+ T\left. \left(\, r_+^3 +l_0^3\left(\, 1 + 3l_0^2/a^2\,\right)/2   \right)\,\right]
\label{fpprimo}
\end{eqnarray}

As previously discussed, $\frac{dF^{Off.}}{dr_+}=0\Rightarrow T=T_H $.

\subsection{``Low'' vacuum pressure phases}

The extrema of free energy indicate existence of both \textit{multiple} and 
\emph{single} regular BHs for different values of the temperature. 
The alternation of single/multiple states is the signature of a first order phase transition,
as in finite temperature quantum field theory, to which we refer.\\ 
It turns out that single/multiple BH transitions occur at the inflection
points of free energy (~extremal points of $T_H$~).
% In order to grasp better what is going on in the near extremal region, we zoom that part of
% Figure(\ref{fasi}) in Figure(\ref{zoom}).\\
Thus,  the following scenario is in place  in the low-pressure regime $a \ge a_c $:\\
\begin{enumerate} 
\item
$T=0$. BH is in the \textit{frozen single state}. 
The only ground state is the extremal configuration with $r_+=r_-=l_0$.
\item $0\le T \le T_{min}$. BHs are in the \textit{cold single} state
of radius $l_0< r_+ < r_{min}$.  This is due to the effect of the minimal length $l_0$. 
\item
At $T=T_{min}$ an inflexion point appears in $F^{off}$ at $r_+=r_{min}$. 
\item
  $T_{min} < T < T^\ast$. New local minimum develops and the system
splits into two  co-existing states. The small
near-extremal BH is energetically favored.  
\item
$T = T^\ast$ the two minima become degenerate
and the system is in a \textit{mixed state}. Both BHs have the same free energy.
\item $T^\ast < T < T_{max}$ large BHs become stable, while near-extremal 
 BHs are only locally stable.
\item $T=T_{max}$ The near-extremal minimum merges with the local maximum.
There is a  new  transition from multiple to a single BH state.  
\item $T>T_{max}$  there is \textit{high temperature, single, stable} BH.
\end{enumerate}
The  above scenario describes  \textit{first order} 
phase transitions from single to multiple BHs at $T=T_{min}$ and $T=T_{max}$.

\begin{figure}[ht!]
\begin{center}
\includegraphics[width=8cm,angle=0]{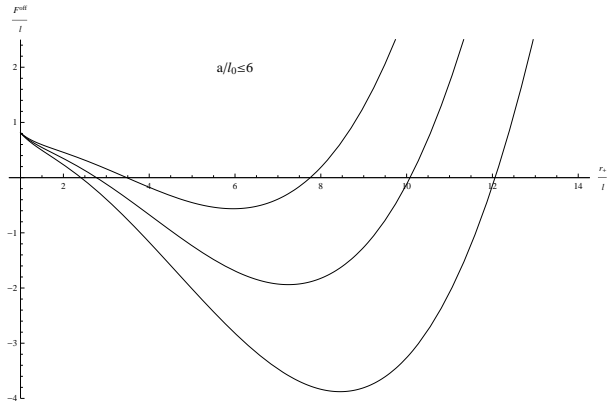}
\caption{\label{fasi2} Plot of the off-shell free energy for different values of the temperature $T$ in the high-pressure regime,
$a/l_0 \le 6$.  }
\end{center}
\end{figure}

\subsection{``High'' vacuum pressure phases}

One can see  that  the first order phase 
transitions take place in the low vacuum pressure regime, only.
 In fact, the two parameters of the theory have 
``opposite'' effects i.e. $l_0 $ dominates short-range distance behavior and lowers $T_{max}$, while $ a $ 
dominates long-range region of $r_+$ and thus raises $T_{min}$.\\
It is reasonable to expect that, at the certain point, these two opposing 
effects will meet creating an \textit{inflexion} point of the temperature.
The confirmation of our conjecture is shown in Fig.(\ref{tempads}). \\
In terms of the free energy the same effect can be seen in the Figure(\ref{fasi2}), where 
a single minimum exists beyond the inflexion point $a=2.88\, l_0 $. 
Varying $T$ only lowers the \textit{position} of the \textit{single} minimum. \\
A complex phase structure is a general feature of several types of BH where some characteristic length
scale is present in the metric \cite{Kubiznak:2012wp,Smailagic:2012cu,Spallucci:2013osa,Spallucci:2013jja}.

%\newpage
\section{Gaussian regularization}
\label{Gauss}
In the first part of this chapter we introduced an ``~ad hoc~'' method of eliminating curvature singularities 
 using ideas usually adopted in quantum field theory, i. e. introducing a suitable short-distance cut-off. However,
there is another approach in QFT achieving the same goal but having a profound physical significance.
For example, UV infinities in Feynman diagrams (beyond the tree-level) can be eliminated by replacing the bare (euclidean)
propagator with an exponentially damped one \cite{Smailagic:2003yb,Smailagic:2003rp,Smailagic:2004yy,Spallucci:2006zj}

\begin{equation}
 \frac{1}{k^2 + m^2}\longrightarrow \frac{e^{-k^2/ 2\Lambda}}{k^2 + m^2}
\label{regprop}
\end{equation}

The  meaning of the above substitution is to replace a divergent two-point Green function (in coordinate space) with a regular
 one solving the inhomogeneous equation

\begin{equation}
 \left(\, \partial^2 + m^2\,\right) G_\Lambda \left(\, x- y\,\right) = \frac{1}{(2\pi \Lambda)^2}\exp\left(\, -\frac{(x-y)^2}{4\Lambda}\,\right)
\label{reggreen}
\end{equation}
  
Equation (\ref{reggreen}) shows that physical particles with propagator (\ref{regprop}) are not matter ``points'', 
but are smeared Gaussian energy distributions. In quantum mechanics Gaussian wave packets represents minimal uncertainty
states, i.e. the closest one can get to a classical particle.\\
The approach of substituting point-like particles with Gaussian matter distributions has been carried out in a number of papers in order 
to describe quantum mechanics and in QFT in \emph{coordinate} non-commutative space(time),  characterized by

\begin{equation}
 \left[\, x^\mu , x^\nu\,\right]= i \theta^{\mu\nu}
\label{comm}
\end{equation} 

where $l_0$ can be related to Lorentz invariant quantity 

\begin{equation}
 l_0^2 \propto \sqrt{\theta^{\mu\nu}\theta_{\mu\nu} }
\end{equation} 

It is already widely accepted that space-time at short distances is no more modeled by a smooth manifold but something completely
different. Our ignorance about the space-time Planckian phase leaves room for different hypothesis, e.g. string, loop,
fractal, non-commutative, foamy phases, etc., all sharing the existence of  a characteristic length scale. The introduction of $l_0$
as the minimal width of a  Gaussian distribution is motivated by non-commutativity of coordinates (\ref{comm}), much like
the non-commutativity of phase space coordinates in QM is characterized by $\hbar$.\\
By accepting this idea, one wonders  how to implement it in the case of gravity. At first glance, one could think of modifying 
the very definition of the  metric tensor to incorporate $l_0$ in the space-time fabric, e.g. replacing the ordinary
product of functions by the star-product. This approach, apart heavily complicating Einstein equations, has the basic flaw
that any perturbative expansion in theta, truncated at a finite order, leads to the loss of non-locality of the
 original theory. The resulting solutions contain all the pathologies of the commutative theory (curvature singularities)
in spite of having introduced $l_0$ from the very beginning.\\
Alternatively, we argued that instead of changing the space-time geometry 
the effects of $l_0$ can be implemented through the matter
source. The line of reasoning is the following. Metric field is a geometrical structure defined over an underlying
manifold. Curvature measures the strength of the metric field, i.e. is the response to the presence of a mass-energy
distribution.   On the
other hand, energy-momentum density determines space-time curvature. Thus, we conclude that in General Relativity
the effects of $l_0$ can be taken into account by keeping the standard form of the Einstein tensor in
the l.h.s. of the field equations and introducing a modified energy-momentum tensor as a source in the r.h.s. \\
 Thus, we choose
the mass density of a static, spherically symmetric, smeared, particle-like gravitational source as \cite{Nicolini:2005vd}

\begin{equation}
 \rho(r)\equiv M \sigma\left(\, r\,\right)
=\left(\frac{3}{l_0}\right)^3 \frac{M}{(4\pi )^{3/2} }\exp\left( -\frac{9 r^2}{4l_0^2}\,\right)
\label{ro}
\end{equation} 
By solving Einstein equations with (\ref{ro}) as a
matter source, we find the line element \footnote{The numerical coefficients have been numerically determined
 to have $r_0= l_0$.
 }

\begin{eqnarray}
 ds^2=&&-\left(\, 1 -\frac{2MG_N}{r}\frac{\gamma(3/2; 9 r^2/4l_0^2)}{\Gamma(3/2)}\,\right)dt^2 \nonumber\\
&& +\left(\, 1 -\frac{2MG_N}{r}\frac{\gamma(3/2; 9r^2/4l_0^2)}{\Gamma(3/2)}\,\right)^{-1} dr^2
       + r^2 d\Omega^2
\label{nostra}
\end{eqnarray} 

where, 

\begin{equation}
 \gamma(3/2; x)\equiv \int_0^x du\, u^{1/2} e^{-u}
\end{equation} 

Strictly speaking, the density (\ref{ro}) is non-vanishing everywhere, even if it quickly drops below
any measurable value at few orders of $l_0$. However,  this may rise the question if a  BH can be formed by
such a smeared distribution. In order to answer this question we evoke the \emph{hoop conjecture} \cite{thorne} and adapt it
to the present situation \footnote{A quantum formulation of the hoop conjecture has been recently proposed in \cite{Casadio:2013uga}.}. 
It means that we define a \emph{mean radius} of the mass distribution as

\begin{equation}
 <r>\equiv 4\pi \int_0^\infty dr r^2 \sigma(r) r=\frac{4}{3\sqrt{\pi}}l_0
\end{equation} 
 
the hoop conjecture assumes that whenever, for a given total mass $M$, 

\begin{equation}
 <r> \le r_H(M)
\end{equation}  

where, $r_H$ is the radius of the eventual Killing horizon determined from (\ref{nostra}), the metric (\ref{nostra})
will describe a BH.

\begin{figure}[ht!]
\begin{center}
\includegraphics[width=8cm,angle=0]{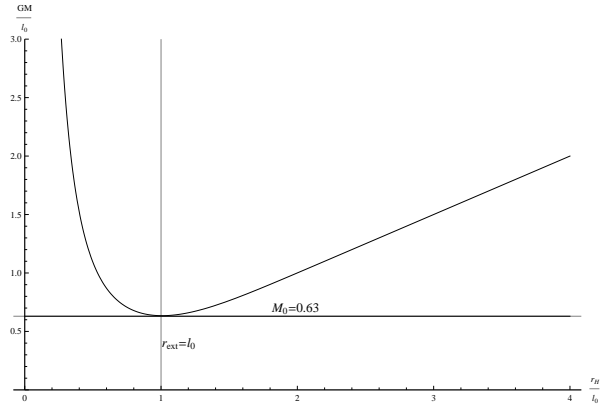}
\caption{\label{qmass} Plot of the mass $M$ as a function of the radius $r_H$ of the horizon. }
\end{center}
\end{figure}

The analysis of the function

\begin{equation}
 M=\frac{r_H}{2G_N}\frac{\Gamma(3/2)}{\gamma(3/2; 9 r^2_H/4l_0^2)}
\end{equation} 

shows (see Fig.()) that  it admits an inner, $r_-$, Cauchy and an outer, $r_+ (\ge r_-)$ Killing horizon, which merge
into a single degenerate horizon, $r_\pm \to r_0$, in the extremal case.  
The radius of the extremal BH is again $r_0=l_0$, as in the toy-model case, while the mass is 

\begin{equation}
 M_0=\frac{l_0}{2G_N}\frac{\Gamma(3/2)}{\gamma(3/2; 9/4)}
\end{equation} 

Thus, we find

\begin{equation}
 <r> < r_0 \le r_+
\end{equation} 

which show that even in the extremal case the mean radius of the mass distribution is smaller that the radius
of the horizon. Enough mass is confined within the horizon in order to sustain the existence of the BH.
\\
Furthermore, the metric (\ref{nostra}) exhibits the short-distance behavior of  the deSitter form with 
no singularity at the origin. This can be inferred from the behavior of the incomplete gamma function 
$ \gamma(3/2; x)$ for small argument

\begin{equation}
 \gamma(3/2; x^2)\approx x^3\ ,\qquad x\longrightarrow 0
\end{equation} 

\subsection{``Bohr quantization''}
A closer look at equation (\ref{ro}) suggests an intriguing analogy with the probability density of the 
the ground-state wave function of an isotropic, 3D, harmonic oscillator

\begin{equation}
\sigma(r)\longleftrightarrow \vert \psi_{000}(r)\vert^2  
\end{equation} 

where, 

\begin{equation}
 \psi_{000}(r) \propto e^{-m r^2\omega/2} 
\end{equation} 

is the ground state wave function. 
In order to identify $m$ and $\omega$ with the corresponding quantities in (\ref{ro}), we notice that 

\begin{equation}
 m\omega =\frac{9}{8l_0^2}
\label{omega}
\end{equation} 

and the mass of the extremal BH represents the minimum energy of the equivalent harmonic oscillator, i.e. $M_0$ is the
zero-point energy

\begin{equation}
 \frac{3}{2}\omega = M_0
\label{emme}
\end{equation}

By solving the two equations (\ref{omega}), (\ref{emme}) we find

\begin{eqnarray}
&& \omega= \frac{2}{3}M_0\ ,\\
&& m= \frac{27}{8}\frac{1}{l_0^2 M_0}
\end{eqnarray}

The non-extremal BH quantized configurations are assumed to be,  the $l=0$ excited states of the corresponding harmonic oscillator:

\begin{equation}
 M_n = M_0 \left(\, 1 + \frac{2}{3} n\,\right) \ ,\quad n=0\ , 2\ , 4\ ,\dots
\label{hspectrum}
\end{equation} 

Due to the spherical symmetry only even oscillator states are allowed. Thus,  in this new formulation of quantum BHs, 
the extremal configuration is pure \emph{collapsed zero-point energy}, and the excited states are non extremal BHs
with discrete masses given by (\ref{hspectrum}). \\
The analogy with the harmonic oscillator \cite{Casadio:2013hja} 
has to be reconciled with the requirement that even excited BH states keep
the simplest geometric structure consisting of  either a single extremal configuration or a  non-extremal BH having
one internal Cauchy horizon and one external Killing horizon only. To achieve this, instead of
using the complete excited harmonic oscillator wave function, 
we propose to keep only the highest power of the Laguerre polynomials in the energy density \footnote{A Maxwell-type energy disitribution
like (\ref{uno}) has been recently used to model the final stage of the gravitational collapse of a ``~thick shell~'' 
of matter \cite{Nicolini:2011fy}. Here we shall use the same type of distribution in a different way.}

\begin{equation}
 \rho_n\left(\, r\,\right) \equiv M_n\sigma_n(r) = \frac{M_n}{\Gamma(n+3/2)}\,
\frac{3^{2n +3} r^{2n} }{2^{2n+4}\pi l_0^{2n+3} }e^{-9r^2/4l_0^2}
 \label{rhon}
\end{equation}

Therefore, the effective geometry describing ``quantized'' BHs is still
described by the equation (\ref{nostra}) with the exception that the continuous parameter $M$ is replaced by its discrete
version given by (\ref{hspectrum})

 \begin{eqnarray}
 ds^2_n=&&-\left(\, 1 -\frac{2M_n G_N}{r}\frac{\gamma(n+3/2\ ; 9 r^2/4l_0^2)}{\Gamma(n+3/2)}\,\right)dt^2 \nonumber\\
&&+ \left(\, 1 -\frac{2M_n G_N}{r}\frac{\gamma(n+3/2\ ; 9r^2/4l_0^2)}{\Gamma(n+3/2)}\,\right)^{-1} dr^2
      + r^2 d\Omega^2
\label{qmetric}
\end{eqnarray} 

For each $n$ the existence of the horizons is given by

\begin{eqnarray}
&& M_n \left(\,r_H\right) = \frac{r_H}{2G_N}\frac{\Gamma(n+3/2)}{\gamma(n+3/2\ ; 9 r^2_H/4l_0^2)}\ ,\\
&& M_n = M_0 \left(\, 1 + \frac{2}{3} n\,\right)\label{mn}
\end{eqnarray} 

However, BH being described as a quantum system one has to verify the average radius of the mass distribution is inside
the corresponding horizon radius. This is known as the  quantum ``~hoop conjecture~''. 

\begin{figure}[ht!]
\begin{center}
\includegraphics[width=8cm,angle=0]{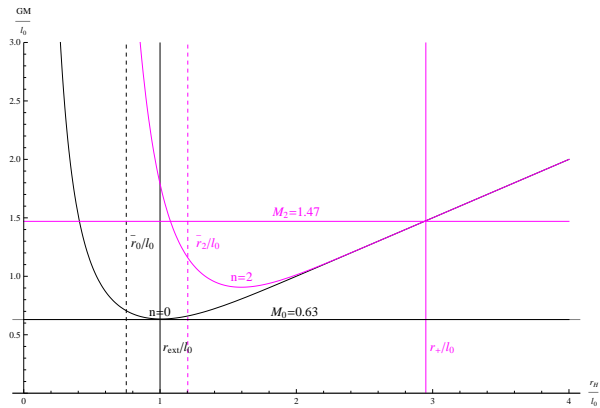}
\caption{\label{spettro} Plot of the ground state $n=0$ (black) and the first excited state $n=2$ (red). Horizontal lines 
correspond to the quantized masses $M_0$, $M_2$. Solid  vertical lines are the Killing horizons. Dashed vertical lines represents
$<r>_n$ of the mass distribution for $n=0\ , 2$. They show as the hoop conjecture is satisfied.}
\end{center}
\end{figure}

\begin{equation}
 < r >_n \equiv 4\pi \int_0^\infty dr r^2 \sigma_n(r) r
\end{equation} 

\begin{equation}
 < r >_n =\frac{2l_0}{3} \frac{\Gamma(n+2) }{\Gamma(n+3/2)} 
\end{equation} 

The figure (\ref{spettro}) below summarizes the main characteristics of the quantized BHs described above. 
The difference between the classical BH given in figure (\ref{mass}). While in figure (\ref{mass}) horizons are obtained
making an arbitrary  cut parallel to the $r_H$-axis, since $M$ is a continuous parameter, in the quantized case we have distinct
graphs for any value of $n$ and for each curve a \emph{single} horizontal line $M=M_n$. Only for $n=0$ there is a an extremal
BH, while for $n\ge 2$ we have non-degenerate BHs corresponding to excited states. For any value of $n$ the hoop conjecture, 
$<r>< r_+$, is satisfied.

\section{Conclusions and discussion}

In this chapter we have described a generalization of the standard Schwarzschild geometry 
which takes into account the presence of a ``minimal'' length $l_0$ which should be present in a quantum, to be, formulation
of gravity.  It is widely believe, from different points of view, that such a parameter should be necessarily incorporated
in a physically meaningful formulation of quantized gravity. In the first part, we have restricted ourself to simple, toy-model,
of regular BH in order to display novel features of such a theory in a relatively transparent way. It has been claimed \cite{Rizzo:2006zb}
that the way in which $l_0$ is introduced is not essential and the same features are common to different models. In fact, we showed
in the second part that a more complicated and realistic model shares basic features with the simpler toy-model. The effects of $l_0$,
in both models,  are: 
\begin{itemize}
 \item to replace the curvature singularity in $r=0$ by a regular deSitter core;
 \item to introduce a lower limit to the BH mass in the form of an extremal configuration, even in the absence of charge and 
       angular momentum;
 \item to identify the radius of the extremal configuration with $l_0$, in agreement with the UV self-complete quantum gravity
       hypothesis, where sub-Planckian distances are screened by (extremal) BH formation;
\item  to provide a consistent final state for BH decay through Hawking radiation in the form of zero temperature extremal BH remnant;
\item  to remove unphysical negative heat capacity during the final stage of BH evaporation; 
\item  to introduce corrections to the area law which is strictly valid for point-like matter sources only.
\end{itemize}

From the technical point of view, $l_0$ was introduced first in the source term in the Einstein equations through the properly
chosen smeared matter distribution. In this picture $l_0$ measures the de-localization, of the otherwise point-like source, induced
by quantum gravitational fluctuations of the underlying space-time manifold. The important advantage of this approach is that 
one can obtain \emph{exact} solutions of the Einstein equations. Alternative approaches, e.g. based on star-product, cannot achieve
the same goal since one is forced to perform truncated perturbative expansion in $l_0$ suffering from the same pathologies $l_0$ is
expected to have cured.\\
The use of a Gaussian energy distribution suggests an intriguing analogy with the ground state of the quantum harmonic oscillator.
This has led us to envisage a possible Bohr-like quantization scheme of our classical regular solution. This has been described in the 
last part of this chapter, where the following quantum picture emerges:

\begin{itemize}
 \item the ground state is an extremal BH of size $l_0$, with mass $M_0$,
 which is the analogue of the harmonic oscillator zero-point energy. 
\item Excited states are non-extremal BHs with mass $M_n= M_0 (1 + 2n/3)$.
\item The semi-classical Hawking picture of thermally radiating BHs remains valid
  for large $n$, while it becomes modified for small $n$. In the truly quantum regime
 the continuous  thermal spectrum turns into a discontinuous spectrum corresponding to the transitions
 between low lying mass levels.
\item  Is it self-consistent to interpret the above excited states as BHs?\\
Our answer is yes, because they satisfy the quantum version of the ``hoop-conjecture'', i.e.
the mean radius of the mass distribution is always smaller that the corresponding radius of the horizon.
\end{itemize}

In view of the quantum picture described above, one can envisage a quantum $BH$ of mass $M_n$ as a ``bound-state''
of $n= 2\ , 4\ , 6\ ,\dots $ quanta of energy $\hbar \omega= 2M_0c^2/3$. The ground-state, or ``\emph{zero-point}'' BH is
given by the extremal configuration with $n=0$ and $M=M_0$.\\
In the last part of this chapter, we were interested in the quantum aspects of regular BHs and have not described the thermodynamical
features of the metric (\ref{nostra}), which however can be found in \cite{Nicolini:2005vd}.

%\newpage

\section{Appendix: regular Coulomb potential}
In Section (\ref{toy}) we regularized the Newtonian potential through the substitution (\ref{uno}). As the Coulomb potential
``suffers'' from the same ``illness'' in $r=0$, one is tempted to regularize it in the same way., i.e. we propose the following 
substitution

 \begin{equation}
  V_C\left(\, r\,\right)\longrightarrow V_{C\, reg.}\left(\, r\,\right) =-\frac{e}{4\pi \varepsilon_0} \frac{r^2}{r^3 + l_0^3/2}
 \end{equation}
The limiting behavior of $V_{C\, reg.}$ is:
\begin{eqnarray}
 && V_{C\, reg.}\left(\, r\,\right)\longrightarrow V_{C}\left(\, r\,\right)\ ,\qquad r\to \infty\ ,\\
 && V_{C\, reg.}\left(\, r\,\right)\longrightarrow -\frac{e}{2\pi \varepsilon_0} \frac{r^2}{ l_0^3} \ ,\qquad r\to 0
\end{eqnarray}

In particular, at short distance the attractive ordinary Coulomb potential turns into a parabolic barrier surrounding the origin. 
Thus, only positive energy, unbound charges can reach the origin.

\begin{equation}
 \frac{dV_{C\, reg.} }{dr}=0\longrightarrow r_{min.}= l_0\ ,\quad V_{C}\left(\, r_{min.}\,\right)
=-\frac{e}{6\pi \varepsilon_0 l_0}
\end{equation}

In the presence of a minimal length, the attractive Coulomb potential develops a minimum at $r=r_{min.}$ and a central
hard core in an analogous manner as the Newtonian potential in the gravitational case.\\
A more physically motivated regularization starts from replacing a point-like charge density with a smeared Gaussian like distribution
given by

\begin{equation}
 \rho(r)
= \frac{e}{(4\pi l_0^2 )^{3/2} }\exp\left( -\frac{r^2}{4l_0^2}\,\right)
\label{roe}
\end{equation} 

Solving the Poisson equation one finds

\begin{equation}
 V_{C\, reg.}= \frac{e}{r}\frac{\gamma(1/2\ ; r^2/4\theta)}{\Gamma(1/2)}
\end{equation} 

and the corresponding electric field, which is the analogue of the gravitational curvature, is regular in $r=0$

\begin{equation}
 E_{C\, reg.}= \frac{e}{r^2}\frac{\gamma(3/2\ ; r^2/4\theta)}{\Gamma(3/2)}
\end{equation} 

This form of the electric field has been used as source in the Einstein equations to obtain the charged extension 
\cite{Ansoldi:2006vg,Spallucci:2009zz} of the BH we discussed in Section (\ref{Gauss}).


\begin{thebibliography}{99}


\bibitem{kiefer}
K.~Kiefer,
``Quantum Gravity'',\ OUP Oxford; (2012)

\bibitem{book1}
A.~Hagar,
``~Discrete or Continuous?: The Quest for Fundamental Length in Modern Physics~''
Cambridge University Press ( 2014)



\bibitem{Fontanini:2005ik} 
  M.~Fontanini, E.~Spallucci and T.~Padmanabhan,
  %``Zero-point length from string fluctuations,''
  Phys.\ Lett.\ B {\bf 633}, 627 (2006)

\bibitem{Aurilia:2013mca} 
  A.~Aurilia and E.~Spallucci,
  %``Why the length of a quantum string cannot be Lorentz contracted,''
  Adv.\ High Energy Phys.\  {\bf 2013}, 531696 (2013)

\bibitem{Smailagic:2010nv} 
  A.~Smailagic and E.~Spallucci,
  %``'Kerrr' black hole: the Lord of the String,''
  Phys.\ Lett.\ B {\bf 688}, 82 (2010)

\bibitem{Nicolini:2008aj} 
  P.~Nicolini,
  %``Noncommutative Black Holes, The Final Appeal To Quantum Gravity: A Review,''
  Int.\ J.\ Mod.\ Phys.\ A {\bf 24}, 1229 (2009)

\bibitem{Hawking:1974sw} 
  S.~W.~Hawking,
  %``Particle Creation by Black Holes,''
  Commun.\ Math.\ Phys.\  {\bf 43}, 199 (1975)

\bibitem{BD}
N.~ D.~ Birrell, P.~C.~W.~ Davies, 
``Quantum Fields in Curved Space''
Cambridge University Press; Reprint edition (1984)

\bibitem{Fulling}
S.~A.~Fulling, 
``Aspects of Quantum Field Theory in Curved Spacetime''
Cambridge University Press (1989)

\bibitem{PD}
L.~Parker and D.~Toms,
``~Quantum Field Theory in Curved Spacetime: Quantized Fields and Gravity~''
Cambridge University Press (2009)

\bibitem{Hollands:2014eia} 
  S.~Hollands and R.~M.~Wald,
  ``Quantum fields in curved spacetime,''
  arXiv:1401.2026 [gr-qc].

\bibitem{Susskind:1994vu}
  L.~Susskind,
  %``The World as a hologram,''
  J.\ Math.\ Phys.\  {\bf 36}, 6377 (1995)

 \bibitem{Susskind:1998dq}
  L.~Susskind and E.~Witten,
  ``The Holographic bound in anti-de Sitter space,''
  arXiv:hep-th/9805114.

\bibitem{Nasseri:2005ji} 
  F.~Nasseri,
  %``Schwarzschild black hole in noncommutative spaces,''
  Gen.\ Rel.\ Grav.\  {\bf 37}, 2223 (2005)

\bibitem{Nasseri:2005yr} 
  F.~Nasseri,
  %``Event horizon of Schwarzschild black hole in noncommutative spaces,''
  Int.\ J.\ Mod.\ Phys.\ D {\bf 15}, 1113 (2006)

\bibitem{Bonanno:2000ep} 
  A.~Bonanno and M.~Reuter,
  %``Renormalization group improved black hole space-times,''
  Phys.\ Rev.\ D {\bf 62}, 043008 (2000)

\bibitem{Hayward:2005gi} 
  S.~A.~Hayward,
  %``Formation and evaporation of regular black holes,''
  Phys.\ Rev.\ Lett.\  {\bf 96}, 031103 (2006)


\bibitem{Spallucci:2012xi} 
  E.~Spallucci and A.~Smailagic,
  %``Black holes production in self-complete quantum gravity,''
  Phys.\ Lett.\ B {\bf 709}, 266 (2012)


\bibitem{Mureika:2011hg} 
  J.~Mureika, P.~Nicolini and E.~Spallucci,
  %``Could any black holes be produced at the LHC?,''
  Phys.\ Rev.\ D {\bf 85}, 106007 (2012)

\bibitem{Nicolini:2013ega} 
  P.~Nicolini, J.~Mureika, E.~Spallucci, E.~Winstanley and M.~Bleicher,
  ``Production and evaporation of Planck scale black holes at the LHC,''
  arXiv:1302.2640 [hep-th].


\bibitem{Spallucci:2011rn} 
  E.~Spallucci and S.~Ansoldi,
  %``Regular black holes in UV self-complete quantum gravity,''
  Phys.\ Lett.\ B {\bf 701}, 471 (2011)


\bibitem{Dvali:2010bf} 
  G.~Dvali and C.~Gomez,
  ``Self-Completeness of Einstein Gravity,''
  arXiv:1005.3497 [hep-th].


\bibitem{Dvali:2010jz} 
  G.~Dvali, G.~F.~Giudice, C.~Gomez and A.~Kehagias,
  %``UV-Completion by Classicalization,''
  JHEP {\bf 1108}, 108 (2011)



\bibitem{Nicolini:2012fy} 
  P.~Nicolini and E.~Spallucci,
  %``Holographic screens in ultraviolet self-complete quantum gravity,''
  Adv.\ High Energy Phys.\  {\bf 2014}, 805684 (2014)



\bibitem{Nicolini:2005vd} 
  P.~Nicolini, A.~Smailagic and E.~Spallucci,
  %``Noncommutative geometry inspired Schwarzschild black hole,''
  Phys.\ Lett.\ B {\bf 632}, 547 (2006)




\bibitem{Maldacena:1997re}
  J.~M.~Maldacena,
  %``The Large N limit of superconformal field theories and supergravity,''
  Adv.\ Theor.\ Math.\ Phys.\  {\bf 2}, 231 (1998)
  [Int.\ J.\ Theor.\ Phys.\  {\bf 38}, 1113 (1999)]
 
\bibitem{Witten:1998zw}
  E.~Witten,
  %``Anti-de Sitter space, thermal phase transition, and confinement in gauge
  %theories,''
  Adv.\ Theor.\ Math.\ Phys.\  {\bf 2}, 505 (1998)
 
 \bibitem{Witten:1998qj}
  E.~Witten,
  %``Anti-de Sitter space and holography,''
  Adv.\ Theor.\ Math.\ Phys.\  {\bf 2}, 253 (1998)
 
 \bibitem{Myers:2008fv}
  R.~C.~Myers, S.~E.~Vazquez,
  %``Quark Soup al dente: Applied Superstring Theory,''
  Class.\ Quant.\ Grav.\  {\bf 25}, 114008 (2008).

 \bibitem{Ambrosetti:2008mt}
  N.~Ambrosetti, J.~Charbonneau and S.~Weinfurtner,
  ``The Fluid/gravity correspondence: Lectures notes from the 2008 Summer
  School on Particles, Fields, and Strings,''
  arXiv:0810.2631 [gr-qc].

  \bibitem{Hartnoll:2009sz}
  S.~A.~Hartnoll,
  %``Lectures on holographic methods for condensed matter physics,''
  Class.\ Quant.\ Grav.\  {\bf 26}, 224002 (2009)

 
\bibitem{Kubiznak:2012wp} 
  D.~Kubiznak and R.~B.~Mann,
  %``P-V criticality of charged AdS black holes,''
  JHEP {\bf 1207}, 033 (2012)

\bibitem{Smailagic:2012cu} 
  A.~Smailagic and E.~Spallucci,
  %``Thermodynamical phases of a regular SAdS black hole,''
  Int.\ J.\ Mod.\ Phys.\ D {\bf 22}, 1350010 (2013)

\bibitem{Spallucci:2013osa} 
  E.~Spallucci and A.~Smailagic,
  %``Maxwell's equal area law for charged Anti-deSitter black holes,''
  Phys.\ Lett.\ B {\bf 723}, 436 (2013)

\bibitem{Spallucci:2013jja} 
  E.~Spallucci and A.~Smailagic,
  %``Maxwell's equal area law and the Hawking-Page phase transition,''
  J.\ Grav.\  {\bf 2013}, 525696 (2013)

\bibitem{Smailagic:2003yb} 
  A.~Smailagic and E.~Spallucci,
  %``Feynman path integral on the noncommutative plane,''
  J.\ Phys.\ A {\bf 36}, L467 (2003)

\bibitem{Smailagic:2003rp} 
  A.~Smailagic and E.~Spallucci,
  %``UV divergence free QFT on noncommutative plane,''
  J.\ Phys.\ A {\bf 36}, L517 (2003)

\bibitem{Smailagic:2004yy} 
  A.~Smailagic and E.~Spallucci,
  %``Lorentz invariance, unitarity in UV-finite of QFT on noncommutative spacetime,''
  J.\ Phys.\ A {\bf 37}, 1 (2004)
  [Erratum-ibid.\ A {\bf 37}, 7169 (2004)]

\bibitem{Spallucci:2006zj} 
  E.~Spallucci, A.~Smailagic and P.~Nicolini,
  %``Trace Anomaly in Quantum Spacetime Manifold,''
  Phys.\ Rev.\ D {\bf 73}, 084004 (2006)



\bibitem{thorne}
  K.~S.~Thorne, {\em Nonspherical gravitational collapse, a short review},
in *J R Klauder, Magic Without Magic*, Freeman, San Francisco 1972, 231-258.

\

\bibitem{Casadio:2013uga} 
  R.~Casadio, O.~Micu and F.~Scardigli,
  %``Quantum hoop conjecture: Black hole formation by particle collisions,''
  Phys.\ Lett.\ B {\bf 732}, 105 (2014)

\bibitem{Casadio:2013hja} 
  R.~Casadio and A.~Orlandi,
  %``Quantum Harmonic Black Holes,''
  JHEP {\bf 1308}, 025 (2013)

\bibitem{Nicolini:2011fy} 
  P.~Nicolini, A.~Orlandi and E.~Spallucci,
  %``The final stage of gravitationally collapsed thick matter layers,''
  Adv.\ High Energy Phys.\  {\bf 2013}, 812084 (2013)


\bibitem{Rizzo:2006zb} 
  T.~G.~Rizzo,
  %``Noncommutative Inspired Black Holes in Extra Dimensions,''
  JHEP {\bf 0609}, 021 (2006)


\bibitem{Ansoldi:2006vg} 
  S.~Ansoldi, P.~Nicolini, A.~Smailagic and E.~Spallucci,
  %``Noncommutative geometry inspired charged black holes,''
  Phys.\ Lett.\ B {\bf 645}, 261 (2007)

\bibitem{Spallucci:2009zz} 
  E.~Spallucci, A.~Smailagic and P.~Nicolini,
  %``Non-commutative geometry inspired higher-dimensional charged black holes,''
  Phys.\ Lett.\ B {\bf 670}, 449 (2009)

\end{thebibliography}
\end{document}